# Quantifying interaction mechanism in infinite layer nickelate superconductors

Evgeny F. Talantsev[1,2]

[1]M.N. Miheev Institute of Metal Physics, Ural Branch, Russian Academy of Sciences, 18, S. Kovalevskoy St., Ekaterinburg, 620108, Russia
[2]NANOTECH Centre, Ural Federal University, 19 Mira St., Ekaterinburg, 620002, Russia

**Abstract**

The connection between the long-range antiferromagnetic order in cuprates and the high-temperature superconductivity is a scientific problem that has yet to be solved after nearly four decades. The properties and difficulties of describing nickelate superconductors are similar to those of cuprates. Recently, Fowlie *et al* (2022 *Nature Physics* **18** 1043) aimed to detect the antiferromagnetic order in $R_{1-x}Sr_xNiO_2$ (R = Nd, Pr, La; x~0, 0.2) films by using the muon spin rotation (μSR) technique. The research group reported the presence of short-range antiferromagnetic order in every nickelates studied. Here, our goal was to prove that this interaction is present in the nickelate films. We did this by analyzing the temperature dependent resistivity, $\rho(T)$, data from the research group. Global $\rho(T)$ data fits to the advanced Bloch- Grüneisen model showed that each of $R_{1-x}Sr_xNiO_2$ compounds can be characterized by a unique power-law exponent, *p* (where *p*=2 for the electron-electron scattering, *p*=3 for the electron-magnon scattering, and *p*=5 for the electron-phonon scattering), and global characteristic temperature, $T_\omega$ (which has the meaning of the Debye temperature at *p*=5). We found that *p*=2.0 in Nd- and Pr-based compounds, and *p*=1.3 for La-based compound. The latter value does not have any interpretation within established theoretical models. We also analyzed $\rho(T)$ data for $Nd_{1-x}Sr_xNiO_2$ ($0.125 \leq x \leq 0.325$) reported by Lee *et al* (*Nature* **619**, 288 (2023)). Our analysis of nickelates led us to conclude that a new theoretical model is needed to describe $\rho(T)$ in materials exhibiting a short-range antiferromagnetic order.

## I. Introduction.

Predicted in 1999 by Anisimov *et al* [1] high-temperature superconductivity in nickelate compounds was discovered by Li *et al* [2] in $Nd_{0.8}Sr_{0.2}NiO_2$ film twenty years later. Experimental discovery [2] sparked theoretical and experimental studies on a variety of $R_{1-x}A_xNiO_2$ (where R is rare earth, and A = Sr,Ca) thin films [3-45] and bulk materials [46]. Recent reviews of the development of this field can be found elsewhere [47-49].

Both infinite layer nickelate and cuprate superconductors face a challenge in understanding how antiferromagnetism and superconductivity are related, even though recent studies have found that nickelates have unique structural features [36,50-53]. Fowlie *et al* [36] aimed to detect the antiferromagnetic order in $R_{1-x}Sr_xNiO_2$ (R = Nd, Pr, La; x~0; 0.2) films using the muon spin rotation (μSR) technique. Nickelate were studied and found to have short-range antiferromagnetic order. It should be noted that in cuprates [53,54], the long-range antiferromagnetic order exists at low charge carrier doping, but it transforms into the short-range order, while the charge carrier doping is increased.

Fowlie *et al* [36] also splatted each of their $R_{1-x}Sr_xNiO_2$ (R = Nd, Pr, La; x~0.2) films in several pieces and measured the normal state temperature-dependent resistivity, $\rho(T)$, in each of these pieces. This research group made the measured $\rho(T)$ dataset freely available online [36]. Here we point out that there is another technique to discover the dominant charge carrier interaction in a conductor by fitting the $\rho(T)$ data with the advanced Bloch-Grüneisen (BG) equation [55-66]:

$$\rho(T) = \frac{1}{\frac{1}{\rho_{sat}} + \frac{1}{\rho_0 + A \times \left(\frac{T}{T_\omega}\right)^p \int_0^{\frac{T_\omega}{T}} \frac{x^p}{(e^x-1)(1-e^{-x})}dx}} \quad (1)$$

where $\rho_{sat}$, $\rho_0$, $T_\omega$, $A$ and $p$ are free fitting parameters. Eq. 1 has the same number of free-fitting parameters as standard Gaussian/Lorenzian function used for data peak fitting with linear dependent background. Parameters $\rho_{sat}$, $\rho_0$, and $A$ are strongly dependent on the density of structural defects and the variation of the nickelate film thickness across the film area.

The power-law exponent, $p$, in Eq. 1 has several integer values associated with particular charge carrier interaction mechanisms [57,60,61]. For instance, $p = 2$ implies that charge carriers in the conductor obey perfect Landau's Fermi liquid phenomenology, $p = 3$ implies electron-magnon phenomenology, and $p = 5$ implies electron-phonon phenomenology [57,60,61]. In this approach, $T_\omega$ (Eq. 1) is the Debye temperature, $T_\omega \equiv T_\theta$, for the case of $p = 5$. For conductors with $p \neq 5$, $T_\omega$ (Eq. 1) represents the characteristic energy scalar for the dominant charge carrier interaction in a given material.

Based on the experimental fact that different pieces of the same $R_{1-x}Sr_xNiO_2$ (R = Nd, Pr, La; x~0.2) films exhibit reasonably different $\rho(T)$ curves [36], to perform independent $\rho(T)$ fit for each piece of the film, and after which to make the averaging of obtained values, is unreliable approach, because of the non-linear character of the primary equation (Eq. 1). Instead, here we utilized a global data fit, where the full set of $\rho(T)$ curves measured for a given sample was fitted to the system of equations:

$$\begin{cases} \rho_{piece,1}(T) = \dfrac{1}{\dfrac{1}{\rho_{sat,piece,1}} + \dfrac{1}{\rho_{0,piece,1} + A_{piece,1} \times \left(\frac{T}{T_\omega}\right)^p \int_0^{\frac{T_\omega}{T}} \frac{x^p}{(e^x - 1)(1 - e^{-x})} dx}} \\ \rho_{piece,2}(T) = \dfrac{1}{\dfrac{1}{\rho_{sat,piece,2}} + \dfrac{1}{\rho_{0,piece,2} + A_{piece,2} \times \left(\frac{T}{T_\omega}\right)^p \int_0^{\frac{T_\omega}{T}} \frac{x^p}{(e^x - 1)(1 - e^{-x})} dx}} \\ \rho_{piece,3}(T) = \dfrac{1}{\dfrac{1}{\rho_{sat,piece,3}} + \dfrac{1}{\rho_{0,piece,3} + A_{piece,3} \times \left(\frac{T}{T_\omega}\right)^p \int_0^{\frac{T_\omega}{T}} \frac{x^p}{(e^x - 1)(1 - e^{-x})} dx}} \\ \quad \cdots \\ \rho_{piece,N}(T) = \dfrac{1}{\dfrac{1}{\rho_{sat,piece,N}} + \dfrac{1}{\rho_{0,piece,N} + A_{piece,N} \times \left(\frac{T}{T_\omega}\right)^p \int_0^{\frac{T_\omega}{T}} \frac{x^p}{(e^x - 1)(1 - e^{-x})} dx}} \end{cases} \quad (2)$$

where the subscripts $piecei$ indicate the piece number designated by Fowlie *et al* [36] in their $\rho(T)$ database, and $p$ and $T_\omega$ are two global fitting parameters that characterize the type and energy scale of the charge carriers interaction in a given $R_{1-x}Sr_xNiO_2$ (R = Nd, Pr, La; x~0.2) film.

Each $\rho_{piece,i}(T)$ dataset is fitted to Eq. 2 by utilizing three local free-fitting parameters, which are $\rho_{sat,piece,i}$, $\rho_{0,piece,i}$, and $A_{piece,i}$, and two global free-fitting parameters, which are $T_\omega$ and $p$. This implies that the global data fit utilizes by about 40% less of free-fitting parameters, than separate fits of each $\rho_{piece,i}(T)$ to Eq. 1.

Based on this, the converging of the global data fit (with established local and global parameters) is more challenging mathematical task in comparison with separate fits. The global data fit is converged, if the global parameter (or the set of global parameters) are indeed a parameter which is unique for whole set of data. For the given case, the number of free-fitting parameters for individual fits (Eq. 1) is 5x$N$, while the global fit exhibit *3x$N$+2* free-fitting parameters, and the advantage to use the global fit is achieved even for *N=2*.

In this study, we found that the fit to Eq. 2 of the full datasets of each $R_{1-x}Sr_xNiO_2$ (R = Nd, Pr, La; x~ 0.2) film converged, and the following parameters were deduced:

1. $p = 2.07 \pm 0.03$ and $T_\omega = 442 \pm 5\ K$ for $Nd_{0.825}Sr_{0.175}NiO_2$ film;
2. $p = 2.00 \pm 0.02$ and $T_\omega = 319.6 \pm 1.5\ K$ for $Pr_{0.80}Sr_{0.20}NiO_2$ film;
3. $p = 1.34 \pm 0.01$ and $T_\omega = 484 \pm 5\ K$ for $La_{0.80}Sr_{0.20}NiO_2$ film.

The deduced *p*-values for Nd- and Pr-based compounds indicate the dominance of the electron-electron interaction in these materials, or, in other words, Fermi liquid behaviour, which is typical for superconducting cuprates at high levels of doping. However, a low deduced $p = 1.34 \pm 0.01$ for $La_{0.80}Sr_{0.20}NiO_2$ film does not have any established theoretical explanation to date. For instance, similarly low *p*-values, $p \sim 1.5$, were deduced by fitting to Eq. 1 $\rho(T)$ data for twisted bilayer graphene (TBG) [66] (raw data reported by Polshin *et al* [67]). However, for TBG, one possible interpretation for low *p*-values can be based on the electron-acoustic phonon interaction model [68], which is unlikely to be applicable for the infinite layer nickelates. From another hand, the closest case for the deduced $p = 1.34 \pm 0.01$ is linear $\rho(T)$ dependence in cuprates at some doping levels (the latter case can be described as $p \to 1.0$ in Eq. 1). Even though the linear $\rho(T)$ dependence in cuprates still does not have a widely accepted description/explanation [69-72], we can propose that the difference in deduced $p$-value originates from the difference in phase diagrams (i.e., doping vs transition temperature) for Nd- and Pr-based compounds and La-based compound.

We also revisited the $\rho(T)$ phase diagram for $Nd_{1-x}Sr_xNiO_2$ ($0.125 \leq x \leq 0.325$) thin films proposed by Lee *et al* [73], based on a single $\rho(T)$ dataset fitted to Eq. 1 vs the full range of doping (i.e. $0.125 \leq x \leq 0.325$).

## II. Raw data source and fitting routine

Before performing a global fit for the $\rho(T)$ data reported by Fowlie *et al* [36], we fitted the single $\rho(T)$ datasets for $Nd_{0.80}Sr_{0.20}NiO_2$ reported by Zhou *et al* [74]. We also fitted $\rho(T)$ data for $Nd_{1-x}Sr_xNiO_2$ ($0.125 \leq x \leq 0.325$) reported by Lee *et al* [73] to Eq. 1.

Raw $\rho(T)$ datasets for global fits for $R_{1-x}Sr_xNiO_2$ (R = Nd, Pr, La; x~0.2) films were obtained from the data source [75] for Supplementary Figure 2 [76] provided by Fowlie *et al* [36]. We retained the same designations for sample pieces as implemented in the data source [75]. Fits were performed using the Levenberg–Marquardt algorithm in the non-linear fitting package of Origin software (ver. 2017, Origin Lab, Northampton, MA, USA).

## III. Results

*3.1. $Nd_{0.80}Sr_{0.20}NiO_2$ film (single dataset)*

Zhou *et al* [74] reported a $\rho(T)$ dataset for $Nd_{0.80}Sr_{0.20}NiO_2$ film deposited on $SrTiO_3$ single crystal in their Fig. 2,a [74]. Zhou *et al* [74] also reported the $\rho(T)$ dataset for $Pt/Co_{0.90}Fe_{0.10}/Nd_{0.80}Sr_{0.20}NiO_2/SrTiO_3$ heterostructure. The nickelate film's parameters cannot be extracted from the total $\rho(T)$ dataset due to the conductive properties of the platinum and $Co_{0.90}Fe_{0.10}$ layers.

The fit of $\rho(T)$ dataset for Nd$_{0.80}$Sr$_{0.20}$NiO$_2$ film deposited on SrTiO$_3$ [74] to Eq. 1 is shown in Fig. 1,a. Deduced parameters are: $p = 2.43 \pm 0.01$ and $T_\omega = 850 \pm 3\ K$. Deduced $p$-value is between two characteristic values of $p = 2.0$ and $p = 3.0$ associated with the electron-electron and the electron-magnon scattering respectively.

Even though the $\rho(T)$ fit to Eq. 1 converged with high quality and deduced $p$ and $T_\omega$ values are within expected values for well-established models (in particular, the Fermi liquid model), there is a remaining question of general interest that can the superconducting state in nickelates be understood within electron-phonon phenomenology. We used the BG model to analyse the $\rho(T)$ (Fig. 1,b) and find the Debye temperature, $T_\theta$, which is a primary factor in understanding electron-phonon mediated superconductivity in nickelates. The value of $p$ was fixed at $p = 5.0$.

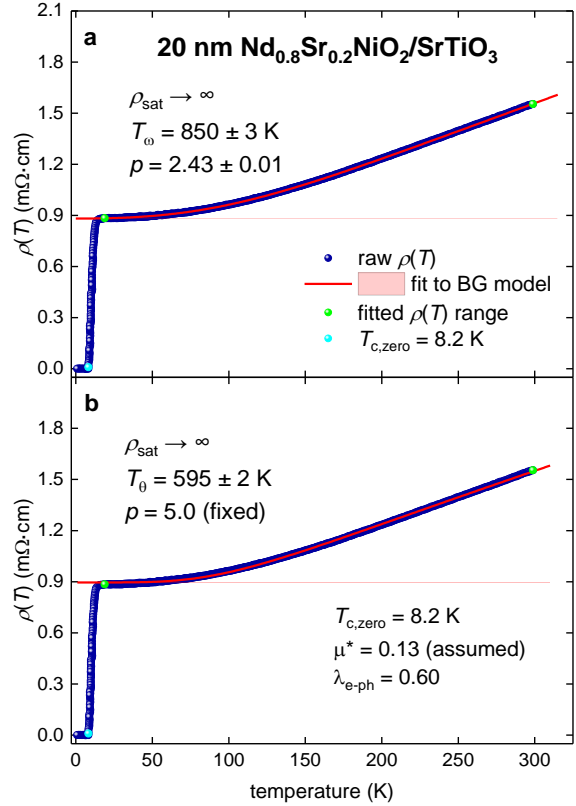

**Figure 1.** $\rho(T)$ data for 20 nm thick Nd$_{0.80}$Sr$_{0.20}$NiO$_2$ film deposited on SrTiO$_3$ single crystal and data fit to Eq. 1 (raw data reported by Zhou *et al* [74]). Green balls indicate the bounds for which $\rho(T)$ data was used for the fit. Cyan ball indicates $T_{c,zero} = 8.2\ K$. (a) $p = 2.43 \pm 0.01$, $T_\omega = 850 \pm 3\ K$, $\rho_{sat} \rightarrow \infty$, fit quality is 0.99998. (b) $p = 5.0\ (fixed)$; $\mu^* = 0.13\ (assumed)$; deduced $T_\theta = 595 \pm 2\ K$, $\lambda_{e-ph} = 0.60$, $\rho_{sat} \rightarrow \infty$, fit quality is 0.9993. The thickness of 95% confidence bands (pink shadow areas) is narrower than the width of the fitting lines.

It should be noted that the conventional approach to determine the Debye temperature, $T_\theta$, from the specific heat measurements, cannot be applied to nickelate films because the sample thickness is ~ 5-20 nm and the film contribution to the total measured heat capacity cannot be extracted. Thus, to the best of our knowledge, the only approach to deduce $T_\theta$ from experimental data is to fit the $\rho(T)$ dataset to Eq. 1 at $p = 5.0$ (fixed). The fit is shown in Fig. 1,b, where deduced $T_\theta = 595 \pm 2\ K$.

The resulting approximate curve for the fit has reasonably high quality (Fig. 1,b), and from the deduced $T_\theta$ value and observed in experiment $T_c$ one can calculate the electron-phonon coupling constant, $\lambda_{e-ph}$, as a root of McMillan-based [77-79] system of equations [65]:

$$T_c = \left(\frac{1}{1.45}\right) \times T_\theta \times e^{-\left(\frac{1.04(1+\lambda_{e-ph})}{\lambda_{e-ph}-\mu^*(1+0.62\lambda_{e-ph})}\right)} \times f_1 \times f_2^* \qquad (3)$$

where

$$f_1 = \left(1 + \left(\frac{\lambda_{e-ph}}{2.46(1+3.8\mu^*)}\right)^{3/2}\right)^{1/3} \qquad (4)$$

$$f_2^* = 1 + (0.0241 - 0.0735 \times \mu^*) \times \lambda_{e-ph}^2. \qquad (5)$$

where $\mu^*$ is the Coulomb pseudopotential parameter, which can be assumed (for the first order of approximation) to be $\mu^* = 0.13$.

As a result, the derived $\lambda_{e-ph} = 0.60$, which is within the weak-coupling range of the electron-phonon mediated phenomenology [80]. It should be noted, that similar to Eq. 3 calculations based on the deduced $T_\theta$ (from thermal capacity measurements) for cuprates give unrealistically high values for $\lambda_{e-ph} \gtrsim 4$ [81]. This fact is a major reason why scientists are studying non-electron-phonon mediated superconductivity in cuprates. However, the deduced value for the Nd$_{0.80}$Sr$_{0.20}$NiO$_2$ film, $\lambda_{e-ph} = 0.60$, can be interpreted as evidence of the possibility (which is not a necessity) of the electron-phonon mediated superconductivity in, at least, the Nd$_{0.80}$Sr$_{0.20}$NiO$_2$ nickelate film.

*3.2. Nd$_{0.825}$Sr$_{0.175}$NiO$_2$ film (four datasets)*

We performed independent fits of the $\rho(T)$ datasets (Fig. 2) for four pieces of Nd$_{0.825}$Sr$_{0.175}$NiO$_2$ film reported by Fowlie *et al* [36] in their Figure S1,a [75]. This was to demonstrate the benefit of the global data fit (Fig. 3) over independent fits (Fig. 2). The film was deposited on a SrTiO$_3$ single crystal. The film thickness is 7.0 nm.

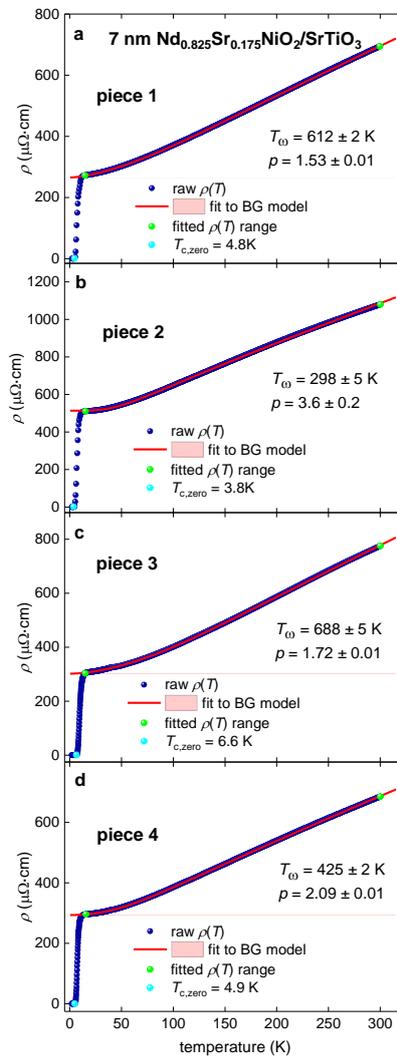

**Figure 2.** $\rho(T)$ for four pieces of the 7 nm thick Nd$_{0.825}$Sr$_{0.175}$NiO$_2$ film deposited on SrTiO$_3$ single crystal and independent data fits to Eq. 1 (raw data reported by Fowlie *et al* [36] in their Figure S1,a [75]). For all fits $\rho_{sat} \to \infty$ (Eq. 1). Green balls indicate the bounds for which $\rho(T)$ data were fitted. Cyan balls indicate $T_{c,zero}$. The goodness of fit: (a) 0.99999; (b) 0.9998; (c) 0.99996; (d) 0.99998. The thickness of 95% confidence bands (pink shadow areas) is narrower than the width of the fitting lines.

It can be seen (Fig. 2) that $T_\omega$ and $p$ are varied within so wide ranges that it is impossible to make any convincing conclusion.

The global data fit, where $T_\omega$ and $p$ are free-fitting parameters, is shown in Fig. 3.

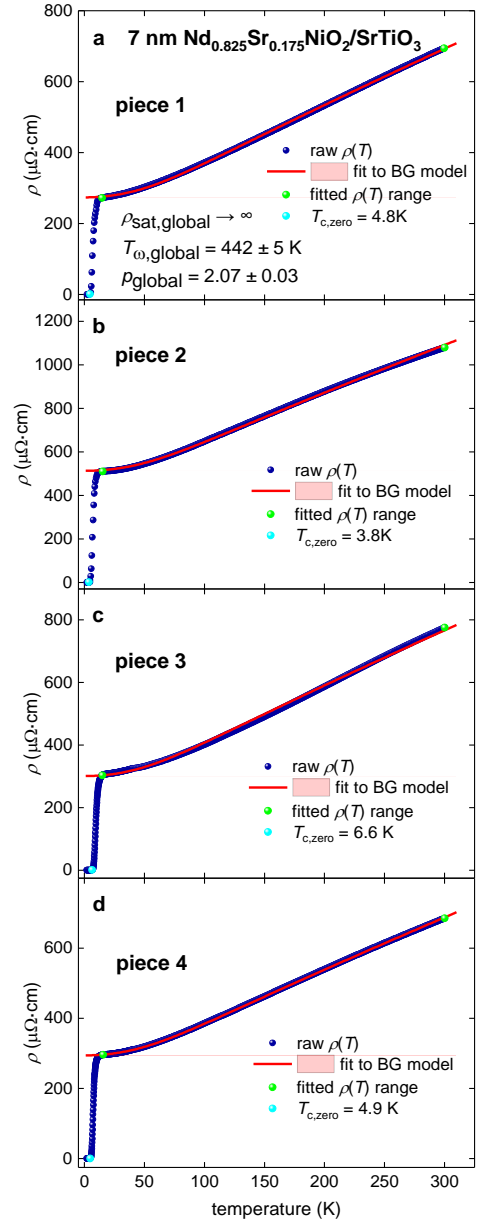

**Figure 3.** $\rho(T)$ data for four pieces of the 7 nm thick $Nd_{0.825}Sr_{0.175}NiO_2$ film deposited on $SrTiO_3$ single crystal and global data fit to Eq. 2 (raw data reported by Fowlie *et al* [36] in their Figure S1,a [75]). For all fits $\rho_{sat} \to \infty$ (Eq. 2). Green balls indicate the bounds for which $\rho(T)$ data were fitted. Cyan balls indicate $T_{c,zero}$. Deduced parameters are $T_\omega = 442 \pm 5\ K$, $p = 2.07 \pm 0.3$. The goodness of fit: (a) 0.9998; (b) 0.9993; (c) 0.9987; (d) 0.99996. The thickness of 95% confidence bands (pink shadow areas) is narrower than the width of the fitting lines.

The deduced free-fitting parameters were: $T_\omega = 442 \pm 5\ K$ and $p = 2.07 \pm 0.03$. The deduced *p*-value is almost the same as the integer characteristic value of $p = 2.0$, which is linked to electron-electron interaction or Fermi-liquid behaviour.

A further step in the analysis was to extract the global Debye temperature, $T_{\theta,global}$, for the same datasets. This implies that global fit was performed at a fixed $p = 5.0$ (see, Fig. 4).

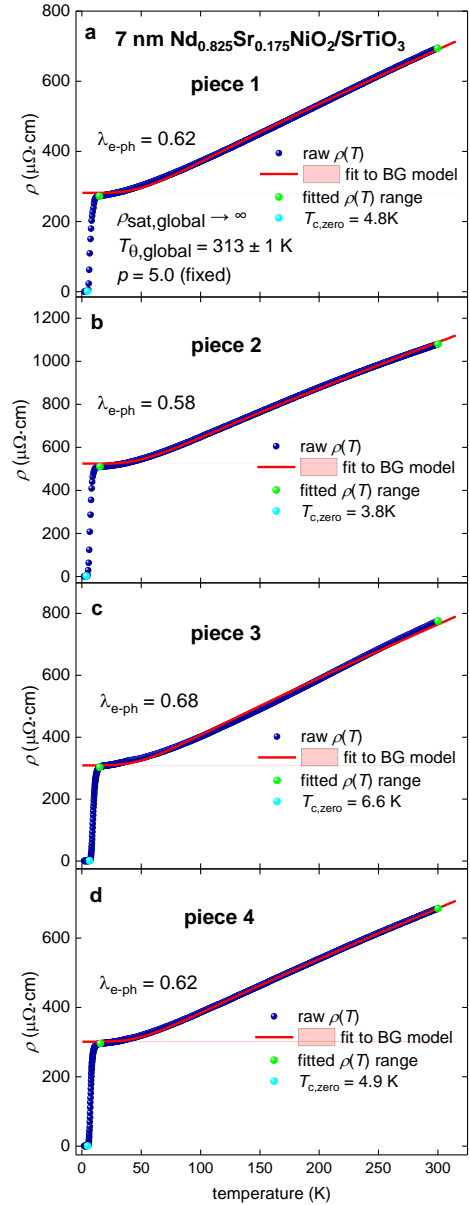

**Figure 4.** $\rho(T)$ data for four pieces of the 7 nm thick $Nd_{0.825}Sr_{0.175}NiO_2$ film deposited on $SrTiO_3$ single crystal and global data fit to Eq. 2 (raw data reported by Fowlie *et al* [36] in their Figure S1,a [75]) at $p = 5.0\ (fixed)$. Green balls indicate the bounds for which $\rho(T)$ data were fitted. Cyan balls indicate $T_{c,zero}$. Deduced Debye temperature is: $T_{\theta,global} = 313 \pm 1\ K$. For all fits $\rho_{sat} \to \infty$ (Eq. 2). The goodness of fit: (a) 0.9992; (b) 0.9995; (c) 0.9981; (d) 0.9997. The thickness of 95% confidence bands (pink shadow areas) is narrower than the width of the fitting lines.

All pieces of the Nd$_{0.825}$Sr$_{0.175}$NiO$_2$ film studied by Fowlie *et al* [36,75] exhibited zero-resistance superconducting transition (when the resistance was reduced below the flicker noise level of the measured electronics and some measured $\rho(T)$ values were read as negative). This implies that the transition temperature, $T_c$, can be defined by the strictest criterion, which is the first negative $\rho(T)$ measurement. However, we used a less strict criterion of $\frac{\rho(T)}{\rho_{norm}} = 0.001$ to define $T_{c,zero}$ for all four pieces of Nd$_{0.825}$Sr$_{0.175}$NiO$_2$.

We calculated the electron-phonon coupling constant $\lambda_{e-ph}$ for each piece using the deduced $T_{\theta,global} = 313 \pm 1\ K$, and the measured $T_{c,zero}$. The calculated values appear to be within a narrow range of $\lambda_{e-ph} = 0.58 - 0.68$ (Fig. 4). These values are in the same ballpark as the $\lambda_{e-ph} = 0.60$ deduced for Nd$_{0.80}$Sr$_{0.20}$NiO$_2$ film (Fig. 1).

Thus, we obtained further evidence that electron-phonon mediated superconductivity can be considered as a possible mechanism to explain the superconductivity in Nd$_{1-x}$Sr$_x$NiO$_2$ films.

*3.3. Pr$_{0.8}$Sr$_{0.2}$NiO$_2$ film (six datasets)*

In Figs. 5,6 we presented another demonstration of the advantage of the global data fit versus independent fits for the case of the Pr$_{0.8}$Sr$_{0.2}$NiO$_2$ film (for which raw $\rho(T)$ datasets were reported by Fowlie *et al* [36] in their Figure S1,b [75]). The film was deposited on a SrTiO$_3$ single crystal. The film thickness is 7.7 nm. The film was divided into six pieces. We used the criterion of $\frac{\rho(T)}{\rho_{norm}} = 0.01$ to define $T_{c,0.01}$ in all six pieces of Pr$_{0.8}$Sr$_{0.2}$NiO$_2$.

In Fig. 5, we performed independent data fitting of the six $\rho(T)$ datasets of the Pr$_{0.8}$Sr$_{0.2}$NiO$_2$ film to Eq. 1, and in Fig. 6, we performed a global fit of the same six $\rho(T)$ datasets to Eq. 2. These fits showed similar results to those found for the Nd$_{0.825}$Sr$_{0.175}$NiO$_2$ film, where the deduced parameters $T_\omega$ and $p$ were within a wide range.

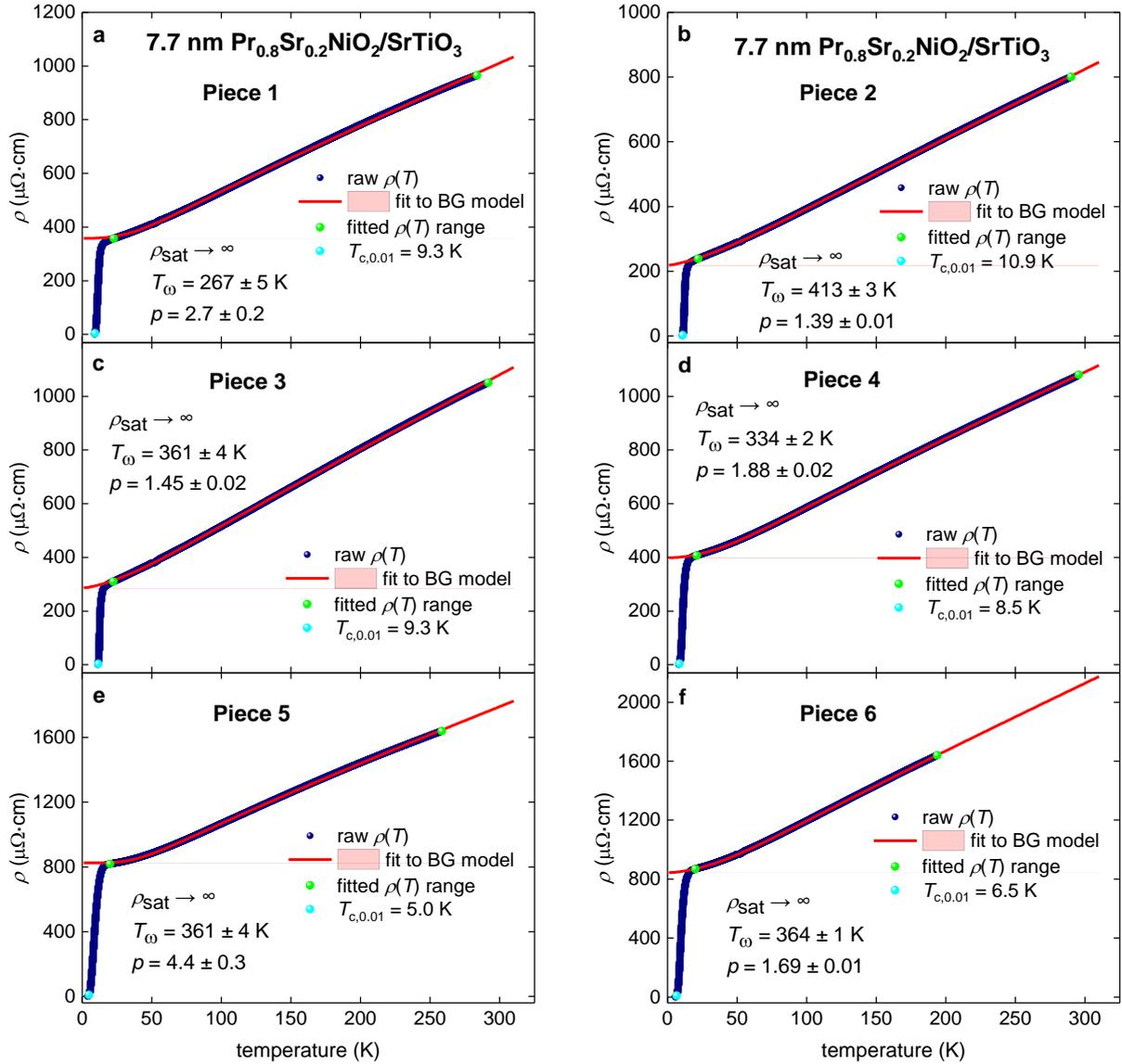

**Figure 5.** $\rho(T)$ for six pieces of the 7.7 nm thick $Pr_{0.8}Sr_{0.2}NiO_2$ film deposited on $SrTiO_3$ single crystal and independent data fits to Eq. 1 (raw data reported by Fowlie *et al* [36] in their Figure S1,b [75]). For all fits $\rho_{sat} \to \infty$ (Eq. 1). Green balls indicate the bounds for which $\rho(T)$ data were fitted. Cyan balls indicate $T_{c,0.01}$. The goodness of fit: (a) 0.9998; (b) 0.99997; (c) 0.99992; (d) 0.99998; (e) 0.9998; (f) 0.99999. The thickness of 95% confidence bands (pink shadow areas) is narrower than the width of the fitting lines.

The global data fit (Fig. 6) shows that the free-fitting parameter $p_{global} = 1.99 \pm 0.02$ is almost the same as the one deduced for $Pr_{0.825}Sr_{0.175}NiO_2$ film (Fig. 4).

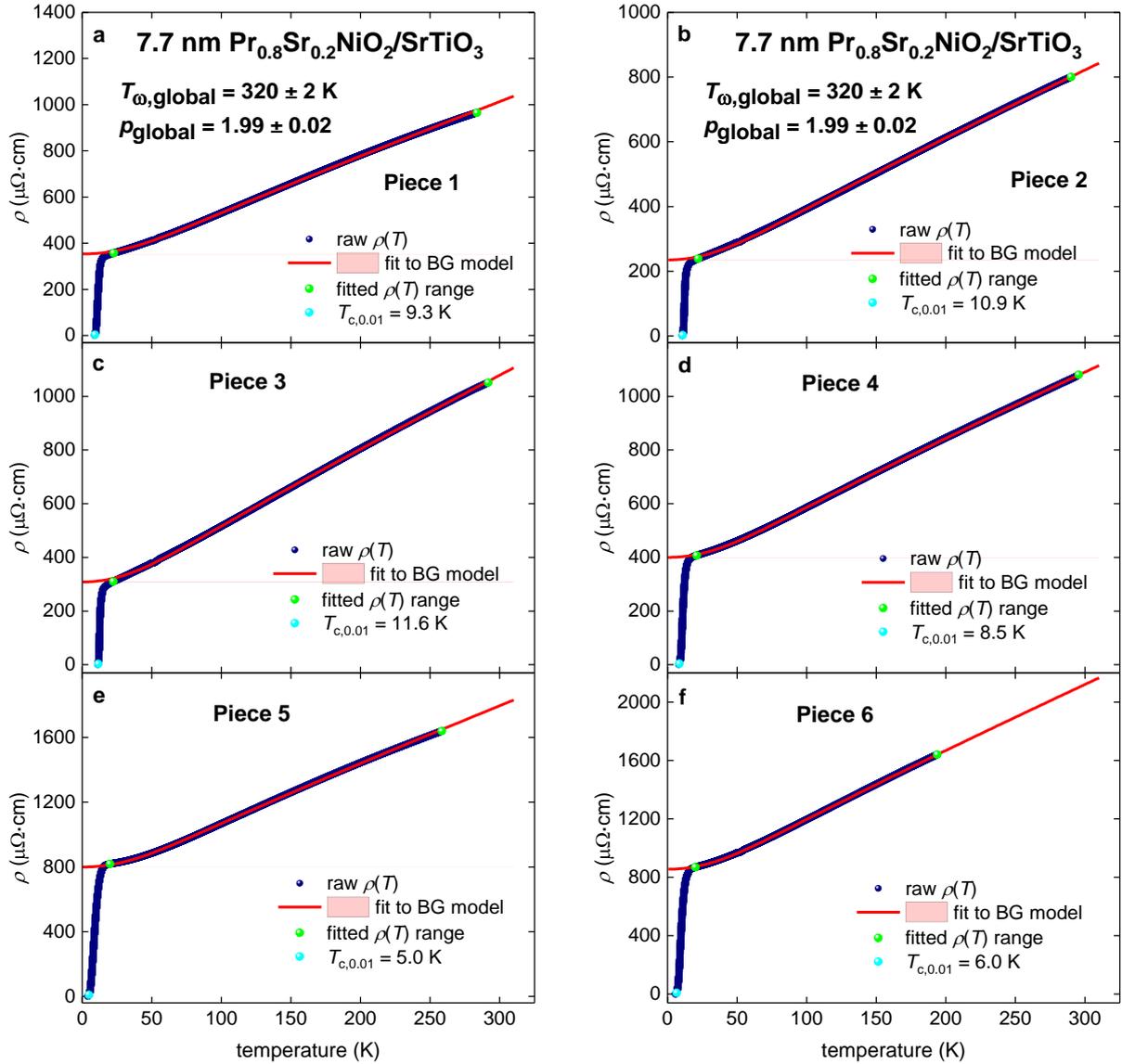

**Figure 6.** $\rho(T)$ for six pieces of the 7.7 nm thick $Pr_{0.8}Sr_{0.2}NiO_2$ film deposited on $SrTiO_3$ single crystal and global data fit to Eq. 2 (raw data reported by Fowlie *et al* [36] in their Figure S1,b [75]). For all fits $\rho_{sat,global} \to \infty$ (Eq. 2). Green balls indicate the bounds for which $\rho(T)$ data were fitted. Cyan balls indicate $T_{c,0.01}$. Deduced $T_{\omega,global} = 320 \pm 2\ K$, $p_{global} = 1.99 \pm 0.02$. The goodness of fit: (a) 0.9997; (b) 0.99990; (c) 0.9998; (d) 0.99998; (e) 0.9998; (f) 0.99996. The thickness of 95% confidence bands (pink shadow areas) is narrower than width of the fitting lines.

The analysis of $\rho(T)$ data within the dominance of the electron-phonon interaction is shown in Fig. 7, where we fitted the same dataset of six $\rho(T)$ curves for $Pr_{0.8}Sr_{0.2}NiO_2$ film to Eq. 2 at $p_{global} = 5.0\ (fixed)$ in Fig. 7.

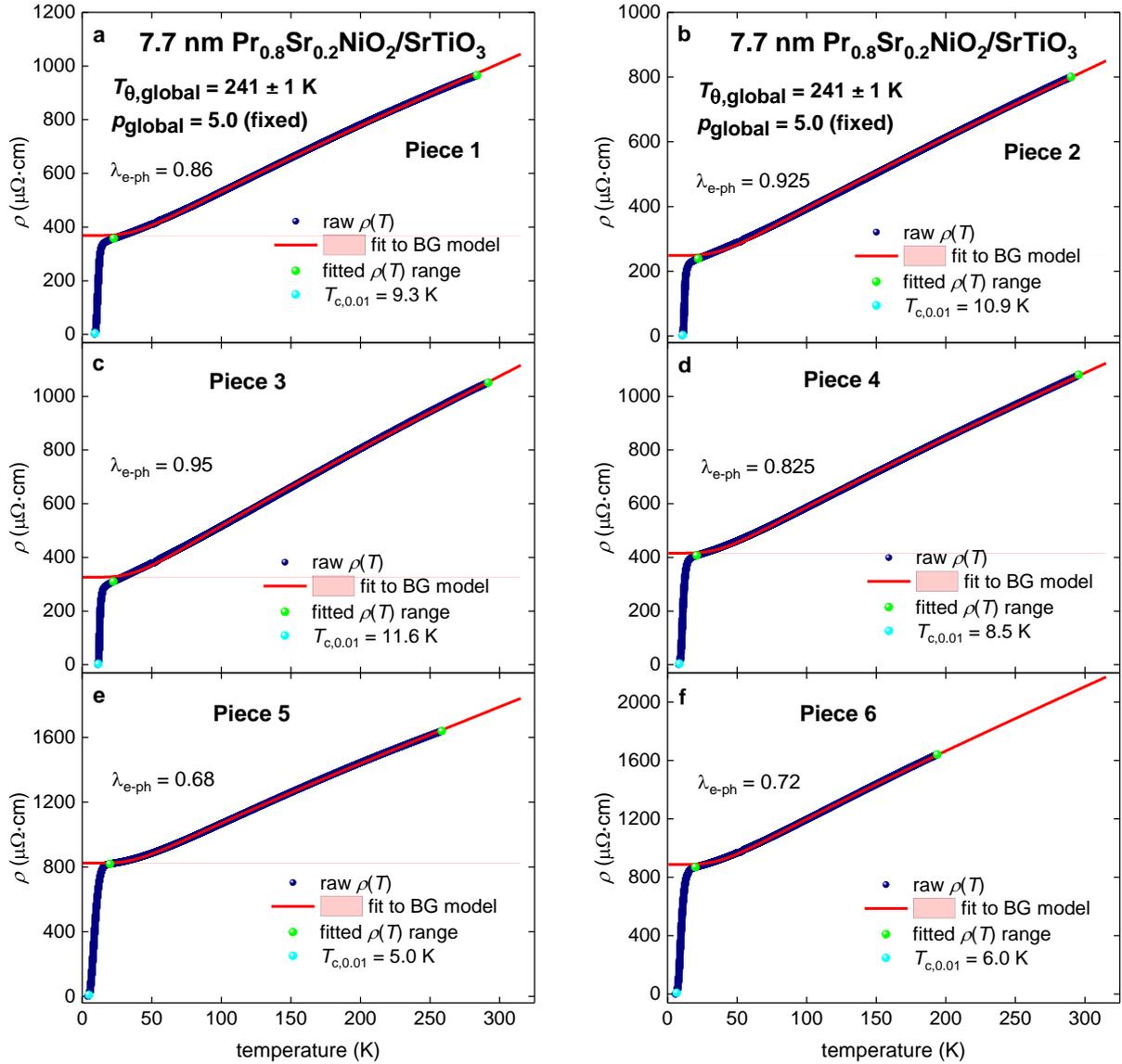

**Figure 7.** $\rho(T)$ data for six pieces of the 7.7 nm thick $Pr_{0.8}Sr_{0.2}NiO_2$ film deposited on $SrTiO_3$ single crystal and global data fit to Eq. 2 (raw data reported by Fowlie *et al* [36] in their Figure S1,a [75]) at $p = 5.0\ (fixed)$. For all fits $\rho_{sat} \to \infty$ (Eq. 2). Green balls indicate the bounds for which $\rho(T)$ data were fitted. Cyan balls indicate $T_{c,zero}$. Deduced Debye temperature is: $T_{\theta,global} = 241 \pm 1\ K$. The goodness of fit: (a) 0.9997; (b) 0.9996; (c) 0.9997; (d) 0.9998; (e) 0.9998; (f) 0.9997. The thickness of 95% confidence bands (pink shadow areas) is narrower than the width of the fitting lines.

We calculated the electron-phonon coupling constant $\lambda_{e-ph}$ for each piece using the measured $T_{c,0.01}$ and deduced $T_{\theta,global} = 241 \pm 1\ K$, The calculated values appear to be within the range of $\lambda_{e-ph} = 0.68 - 0.95$ (Fig. 7), which can be interpreted as moderately strongly coupled values, similar to elemental tin and niobium.

Thus, we obtained further evidence that electron-phonon mediated superconductivity can be considered as a possible mechanism for the superconducting state in Pr$_{1-x}$Sr$_x$NiO$_2$ films.

*3.4. La$_{0.8}$Sr$_{0.2}$NiO$_2$ film (eight datasets)*

For 7.7 nm thick La$_{0.8}$Sr$_{0.2}$NiO$_2$ film deposited on a SrTiO$_3$ single crystal, Fowlie *et al* [36] reported eight unique $\rho(T)$ datasets in their Figure S1,c [75]. In Fig. 8, we showed the independent data fits to Eq. 1 for these eight $\rho(T)$ datasets. It can be observed that the independently deduced free-fitting parameter $p$ varies within a narrow range of $1.29 \leq p \leq 1.54$.

Global data fit at $T_{\omega,global}$ and $p_{global}$ as free-fitting parameters are shown in Fig. 9. Deduced free-fitting parameters are: $T_{\omega,global} = 484 \pm 5\ K$ and $p_{global} = 1.34 \pm 0.01$.

Our attempt to perform global data fit at $p_{global} = 5.0\ (fixed)$ (to deduce the Debye temperature, $T_{\theta,global}$, for La$_{0.8}$Sr$_{0.2}$NiO$_2$ film) failed because the fit diverged. For the same reason, all attempts to perform independent data fit of eight $\rho(T)$ datasets to the model at $p = 5.0\ (fixed)$ also failed.

*3.5. Nd$_{1-x}$Sr$_x$NiO$_2$ film (0.125 ≤ x ≤ 0.325) on LSAT*

Lee *et al* [73] fabricated nearly defect-free Nd$_{1-x}$Sr$_x$NiO$_2$ film (0.05 ≤ x ≤ 0.325) by depositing films on (LaAlO$_3$)$_{0.3}$(Sr$_2$TaAlO$_6$)$_{0.7}$ (LSAT) single crystals. Lee *et al* [73] created the $\rho(T)$ phase diagram for 0.05 ≤ x ≤ 0.325 within traditional linear-parabolic phenomenology [67,68,73,82-84]. Although this simplistic phenomenology is very popular and is applied to a variety of conductors [67,68,73,82-84], it is not necessarily correct, because it fails even for simple cases of pure metals [85].

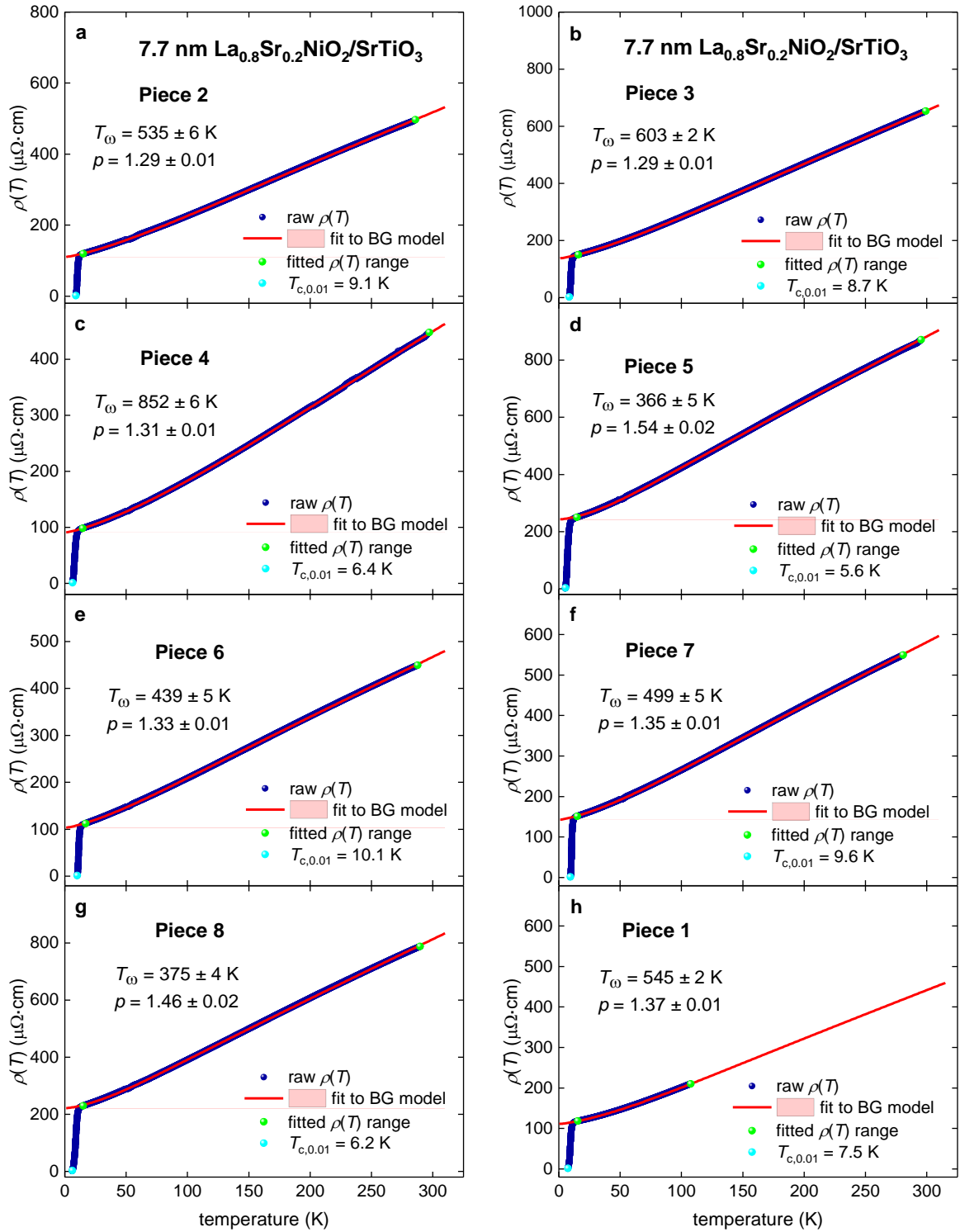

**Figure 8.** $\rho(T)$ for eight pieces of the 7.7 nm thick $La_{0.8}Sr_{0.2}NiO_2$ film deposited on $SrTiO_3$ single crystal and independent data fits to Eq. 1 (raw data reported by Fowlie *et al* [36] in their Figure S1,c [75]). For all fits $\rho_{sat} \to \infty$ (Eq. 1). Green balls indicate the bounds for which $\rho(T)$ data were fitted. Cyan balls indicate $T_{c,0.01}$. The goodness of fit: (a) 0.99995; (b) 1.0; (c) 0.99998; (d) 0.99991; (e) 0.99995; (f) 0.99997; (g) 0.99994; (h) 1.0. The thickness of 95% confidence bands (pink shadow areas) is narrower than the fitting lines width.

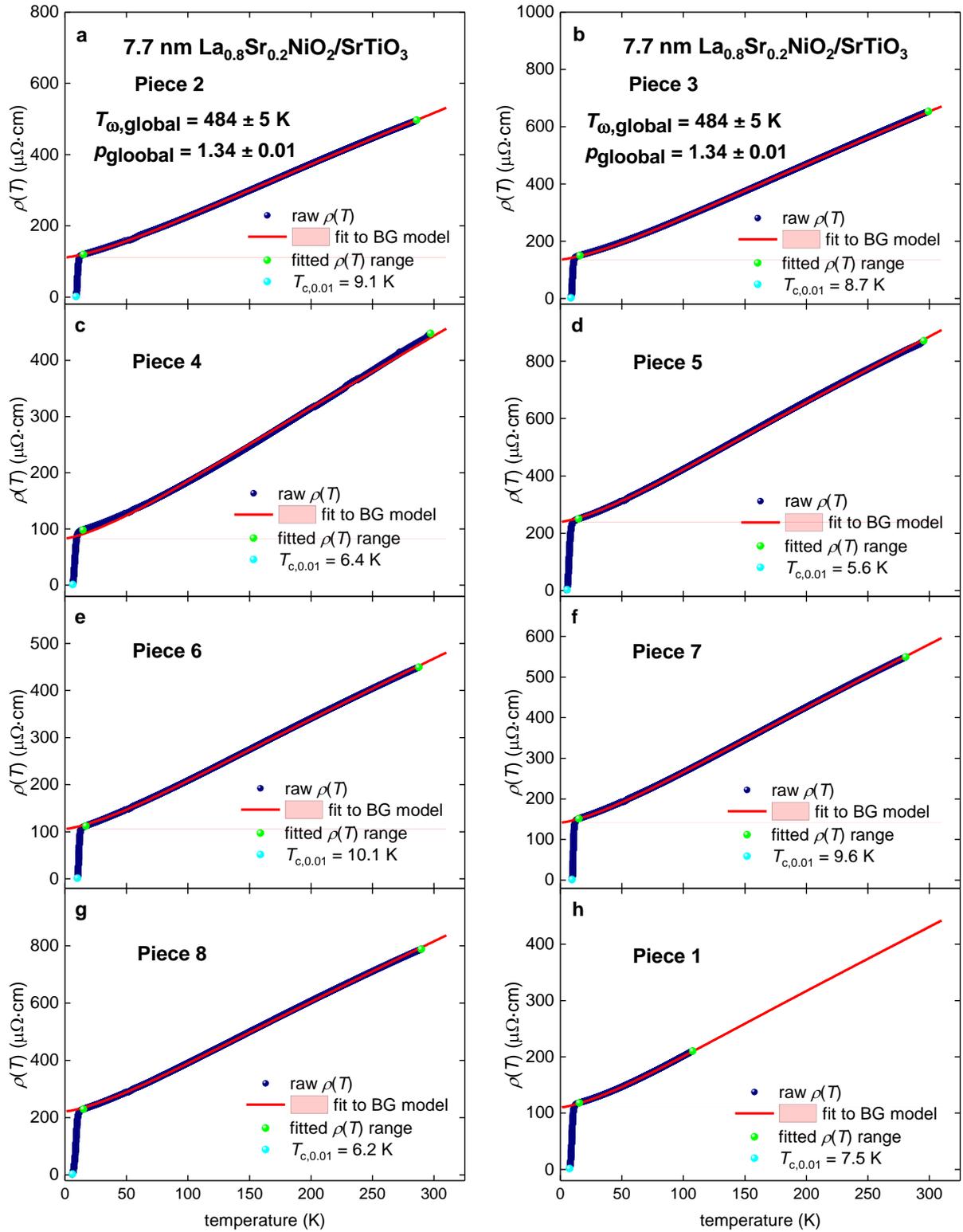

**Figure 9.** $\rho(T)$ for eight pieces of the 7.7 nm thick $La_{0.8}Sr_{0.2}NiO_2$ film deposited on $SrTiO_3$ single crystal and global data fits to Eq. 2 (raw data reported by Fowlie *et al* [36] in their Figure S1,c [75]). For all fits $\rho_{sat} \to \infty$ (Eq. 1). Green balls indicate the bounds for which $\rho(T)$ data were fitted. Cyan balls indicate $T_{c,0.01}$. The goodness of fit: (a) 0.99994; (b) 0.99992; (c) 0.9989; (d) 0.9998; (e) 0.99991; (f) 0.99996; (g) 0.99998; (h) 0.99992. The thickness of 95% confidence bands (pink shadow areas) is narrower than the fitting lines width.

Here, we fitted the $\rho(T)$ datasets for $Nd_{1-x}Sr_xNiO_2$ films ($0.125 \leq x \leq 0.325$) deposited on LSAT (reported by Lee *et al* [73] in their Extended Data Figure 2 [73]) to Eq. 1 (Fig. 10).

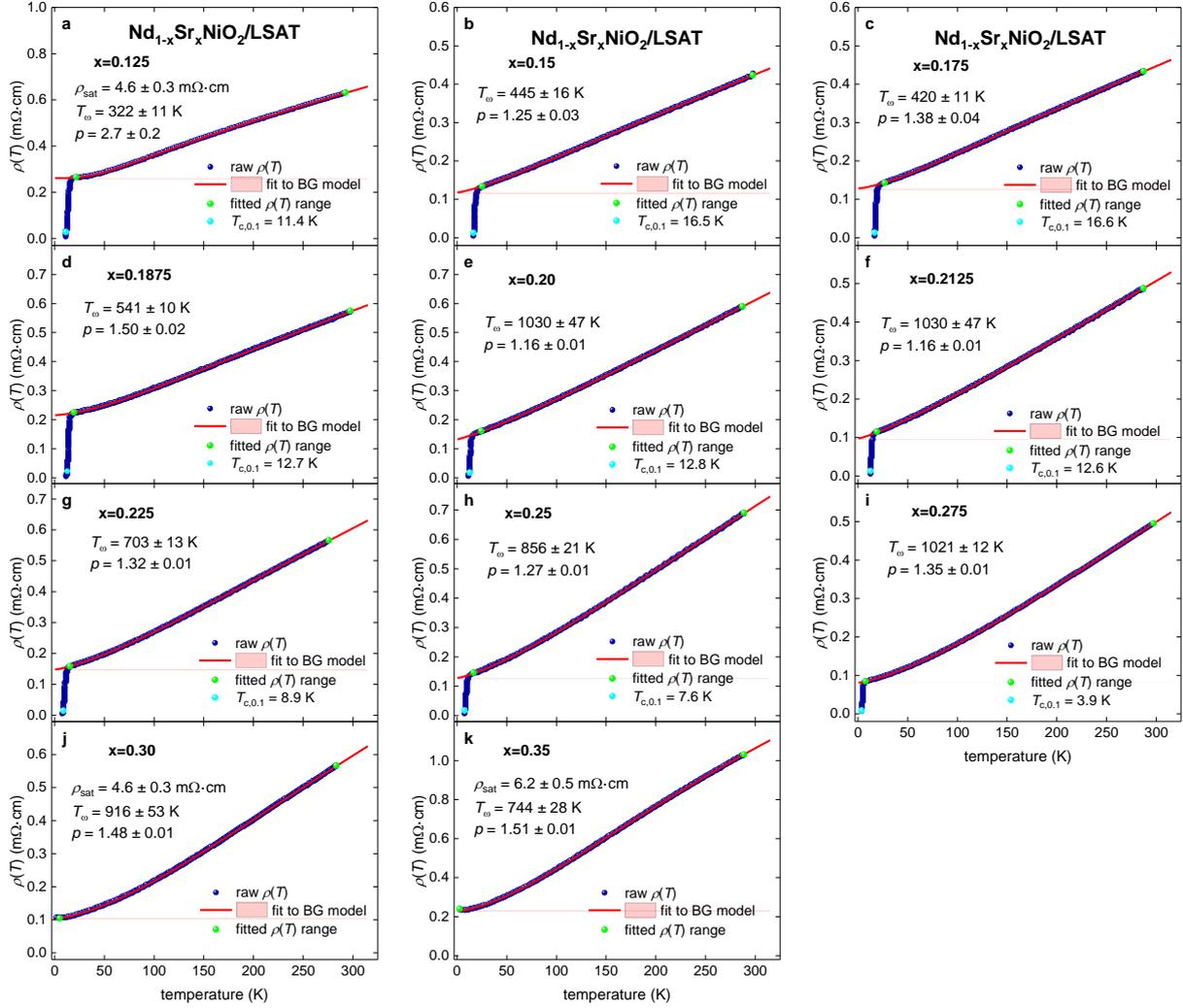

**Figure 10.** $\rho(T)$ for $Nd_{1-x}Sr_xNiO_2$ ($0.125 \leq x \leq 0.325$) films deposited on LSAT single crystal and data fits to Eq. 1 (raw data reported by Lee *et al* [73] in their Extended Data Figure 2 [73]). For all fits, if it is not specified, $\rho_{sat} \to \infty$ (Eq. 1). Green balls indicate the bounds for which $\rho(T)$ data were fitted. Cyan balls indicate $T_{c,0.1}$. The goodness of fit: (a) 0.99995; (b) 0.9998; (c) 0.9998; (d) 0.9998; (e) 0.99992; (f) 0.99992; (g) 0.99992; (h) 0.99992; (i) 0.99998; (j) 0.99994; (j) 0.99996. The thickness of 95% confidence bands (pink shadow areas) is narrower than the fitting lines width.

In Fig. 11, we present deduced parameters as functions of Sr-doping. It can be seen in Fig. 11 that the power-law exponent, $p$, varies within a narrow range of $1.18 \leq p \leq 1.51$ for Sr-doping of $0.15 \leq x \leq 0.325$. This implies that there are no neither parabolic nor linear temperature dependences as it was proposed by Lee *et al*. [73].

It should be noted that at $x = 0.125$ the deduced $p = 2.7 \pm 0.2$, which is only value we were able to deduce in this work, which appears to be close to $p = 3$ (the value attributed to the electron-magnon scattering). This result is different from that reported by Fowlie *et al* [36], who reported the existence of the local antiferromagnetic order in $Nd_{1-x}Sr_xNiO_2$ (x =0.175).

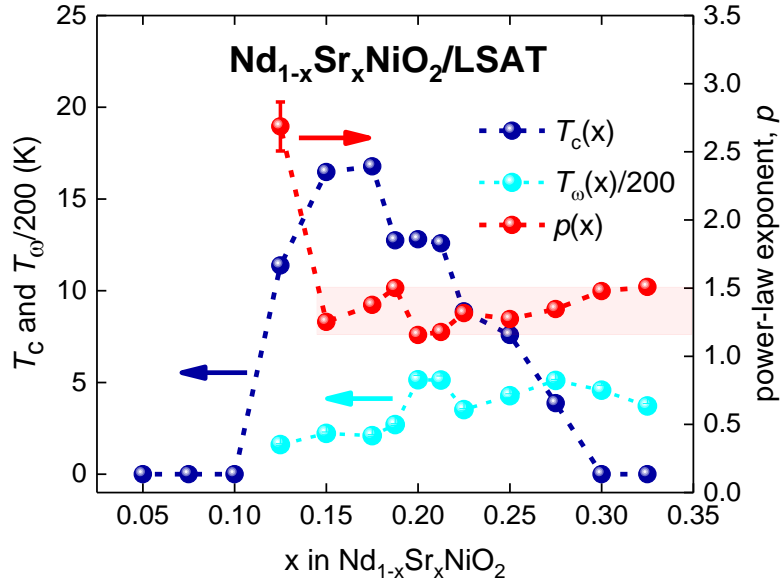

**Figure 11.** Deduced parameters for $Nd_{1-x}Sr_xNiO_2$ (0.125 ≤ x ≤ 0.325) films deposited on LSAT single crystal (raw data reported by Lee *et al* [73] in their Extended Data Figure 2 [73]).

Our result agrees with the phase diagram for cuprates and pnictides, where antiferromagnetic order is exhibited at the low doping state of the phase diagrams.

Here we need to discuss the difference between the deduced parameters for $Nd_{0.80}Sr_{0.20}NiO_2$ films deposited on $SrTiO_3$ (Fig. 1) and LSAT (Fig. 10,e). First, we need to note that the superconducting transition temperature for these films are different, i.e. $T_{c,zero} = 8.2$ K for film on $SrTiO_3$ [74] (Fig. 1) and $T_{c,zero} = 10.5$ K for film on LSAT [73,75]. These films also have different thickness (~ 20 nm for the film on $SrTiO_3$ and ~ 5 nm for the film on LSAT), and there is about one order of magnitude difference in the absolute values of the $\rho(T)$ for these films. The dependence of the $T_c$ and other parameters of thin superconducting films from substrates are well-known experimental fact, where the most prominent dependence is documented for thin films of iron-based superconductors [85]. There is generally accepted understanding that the lattice parameters mismatch between the film and the substrate is the primary origin for the differences in $T_c$ and other parameters of thin film superconductors [85].

Lee *et al* [73,75] extended this phenomenology on nickelate films and showed that the lattice parameters mismatch for $NdNiO_2$ film on $SrTiO_3$ is -0.4%, while for $NdNiO_2$ film on LSAT the mismatch is -1.3 %. Considering that the $Nd_{0.80}Sr_{0.20}NiO_2$ film on LSAT [73,75] is ~4 times thinner than the one deposited on $SrTiO_3$ [74], we proposed to consider that the series of the $Nd_{1-x}Sr_xNiO_2$ ($0.125 \leq x \leq 0.325$) films on LSAT [73,75] exhibit the strong influence of the mechanical strain and also the phononic/polaronic/etc. excitations originated from the substrate.

## IV. Discussion

For more than three decades, primary experimental challenge [2] to synthesise the superconducting nickelates was not a high density of structural defects in the baseline $RNiO_2$ compound, but the unknown routine to synthesize a nickelate with the highly unusual valence of nickel ion, which is $Ni^+$. It should be stressed, that the nickel ions in the most stable nickelates have the valence of $Ni^{2+}$ (which realizes in NiO and $La_2NiO_4$ compounds). Another known valence of nickel is $Ni^{3+}$, which, for instance, is realized in $NdNiO_3$ and $LaNiO_3$ compound.

In the pioneering work by Le *et al* [2], the $NdNiO_3$ epitaxial thin film deposited on $SrTiO_3$ (001) substrate was converted in the $NdNiO_2$ under the strong reduction agent, $CaH_2$ (details can be found in Ref. 2). In the result, the $NdNiO_2$ film remains its epitaxial crystallinity to the $SrTiO_3$ (001) substrate, and after partial Nd substitution by Sr, the $Nd_{1-}$ $_x$Sr$_x$NiO$_2$ films showed the superconducting transition at $T_c \lesssim 15\ K$. This approach was implemented in all following studies for nickelates thin films.

Despite some defects are created during the reduction process, i.e. R$_{1-x}$A$_x$NiO$_3$ → R$_{1-x}$A$_x$NiO$_2$, this structural transformation and the density of the created defects are not dramatic changes in the films crystallinity in comparison with other fabrication technology for thin film superconductors. For instance, we can mention the metalorganic approach for the fabrication of epitaxial RBa$_2$Cu$_3$O$_{7-\delta}$ films [86] which exhibit record high self-field critical current densities, $J_c$(sf,$T$) [86,87], and epitaxial crystallinity alignment to the substrate [86,88,89], despite multiple chemical reactions occur in the solid metalorganic precursor during the RBa$_2$Cu$_3$O$_{7-\delta}$ film synthesis [86]:

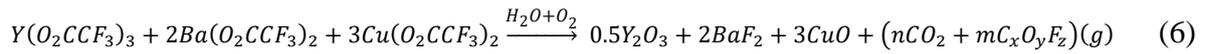

$$Y(O_2CCF_3)_3 + 2Ba(O_2CCF_3)_2 + 3Cu(O_2CCF_3)_2 \xrightarrow{H_2O+O_2} 0.5Y_2O_3 + 2BaF_2 + 3CuO + (nCO_2 + mC_xO_yF_z)(g) \quad (6)$$

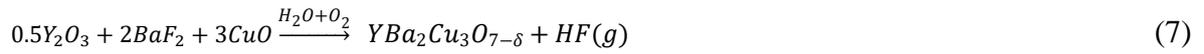

$$0.5Y_2O_3 + 2BaF_2 + 3CuO \xrightarrow{H_2O+O_2} YBa_2Cu_3O_{7-\delta} + HF(g) \quad (7)$$

In the result, despite a large variation in final atomic structure and microscopic defect density in the RBa$_2$Cu$_3$O$_{7-\delta}$ films [88-90] fabricated by different routines, primary superconducting parameters [89,91] (for instance, $T_c$, the London penetration depth, $\lambda(0)$, $J_c$(sf,$T$), etc.) are varying with narrow ranges.

Considering the above, it seems that it is unlikely, that the variation of the Ruddlesden-Popper-type vertical stacking faults density in Nd$_{1-x}$Sr$_x$NiO$_2$ films (reported by Lee et al [73]) can significantly change the $\rho(T)$ temperature dependence. There is no doubts, that these defects determine $\rho_{sat}$, $\rho_0$, and $A$ parameters in Eqs. 1,2, because these parameters determine the absolute value of the $\rho(T)$. However, in our work we derived the $T_\theta$, $T_\omega$ and $p$, which determine the shape of the $\rho(T)$ curve, and these parameters do not determine the absolute $\rho(T)$ values. For instance, in Figures 1-9 the $\rho_{sat}$ is varying within a very wide range:

$$4\ m\Omega \times cm \lesssim \rho_{sat} \leq \infty\ \Omega \times cm \quad (8)$$

while $T_\theta$, $T_\omega$ and $p$ are varying within a range observed for other conductors.

Another important issue is that Eq. 1 was applied for thousands of conductors which all were not defect-free samples. It looks very unlikely, that the Ruddlesden-Popper-type vertical stacking faults or other type of planar defects can modify the $\rho(T)$ curves in nickelates that Eq. 1 is no any longer valid tool to analyse the $\rho(T)$.

We also need to note that our primary Eq. 1 has the same number of free-fitting parameters as standard Gaussian/Lorenzian function used for data peak fitting with linearly dependent background. Parameters $\rho_{sat}$, $\rho_0$, and $A$ are strongly dependent on the density of structural defects and the variation of the film thickness across the film area (which might be different from nominal film thickness). However, the $T_\theta$, $T_\omega$ and $p$ (which describe the strength and type of charge carriers interaction in the solid) are very unlikely to be dependent from the defect density until epitaxial crystallinity remains for the nickelate films.

## V. Conclusions

The main result of our study is that the measurements of a single dataset (for instance, $\rho(T)$ curve), which is a conventional way to conduct experimental studies for a thin film of materials with strongly correlated charge carriers [82-84,92-98], is not sufficient to perform convincing analysis and conclusions. Despite the deduced parameters for independent data fits for $Nd_{0.825}Sr_{0.175}NiO_2$ and $Pr_{0.8}Sr_{0.2}NiO_2$ films exhibiting large variations, the global data fits revealed expected, for these films, the Fermi-liquid behaviour of the $\rho(T)$ curves. This is perhaps the most prominent demonstration of the advantage of global data analysis over single or independent data fit.

Our analysis shows that assuming all nickelates have the same charge carrier interaction mechanism is not entirely accurate. For instance, we found that rare earth element, R, plays the dominant role in the charge carriers interaction in $R_{1-x}Sr_xNiO_2$ (R = Nd, Pr, La; x~0, 0.2) thin films. As a direct consequence of this, it is incorrect to assume that all nickelates exhibit the same pairing mechanism.

We also explored the possibility that the nickelates exhibit the electron-phonon mediated superconductivity. To do this, we extracted the Debye temperature, $T_\theta$, from experimental $\rho(T)$ curves. Deduced $T_\theta = 240 - 600\ K$ for $Nd_{0.825}Sr_{0.175}NiO_2/SrTiO_3$ and $Pr_{0.8}Sr_{0.2}NiO_2/SrTiO_3$ films are high enough that the possibility for the electron-phonon mechanism with $T_c \lesssim 15\ K$ should not be rejected from the consideration. We call for much deeper understanding of the electron-phonon interaction in nickeltaes, because this interaction *de-facto* is excluded from current theoretical studies, while several unconventional and exotic pairing mechanisms have been proposed.

Following recent study by Fowlie et al [75] on μSR studies [75] of nickelates which showed the presence of a local antiferromagnetic order in $R_{1-x}Sr_xNiO_2$ (R = Nd, Pr, La; x∼0, 0.2) [75]. We analyse of $\rho(T)$ data measured for same films and show that some deduced parameters are in the range that does not cover by any existing theory. We call for further theoretical development of the $\rho(T)$ data interpretation.

**Acknowledgment**

The author thanks Jennifer Fowlie (Stanford University) and all co-workers of Ref. 75 for making raw experimental data freely available, which makes it possible to perform this study. The author thanks the financial support provided by the Ministry of Science and Higher Education of Russia (theme "Pressure" No. 122021000032-5). The research funding from the Ministry of Science and Higher Education of the Russian Federation (Ural Federal University Program of Development within the Priority-2030 Program) is gratefully acknowledged.

## Data availability statement

The data that support the findings of this study are available from the corresponding author upon reasonable request.

## Declaration of interests

The author declares that he has no known competing financial interests or personal relationships that could have appeared to influence the work reported in this paper.

**Figures Captions**

**Figure 1.** $\rho(T)$ data for 20 nm thick $Nd_{0.80}Sr_{0.20}NiO_2$ film deposited on $SrTiO_3$ single crystal and data fit to Eq. 1 (raw data reported by Zhou *et al* [74]). Green balls indicate the bounds for which $\rho(T)$ data was used for the fit. Cyan ball indicates $T_{c,zero} = 8.2\ K$. (a) $p = 2.43 \pm 0.01$, $T_\omega = 850 \pm 3\ K$, $\rho_{sat} \to \infty$, fit quality is 0.99998. (b) $p = 5.0\ (fixed)$; $\mu^* = 0.13\ (assumed)$; deduced $T_\theta = 595 \pm 2\ K$, $\lambda_{e-ph} = 0.60$, $\rho_{sat} \to \infty$, fit quality is 0.9993. The thickness of 95% confidence bands (pink shadow areas) is narrower than the width of the fitting lines.

**Figure 2.** $\rho(T)$ for four pieces of the 7 nm thick $Nd_{0.825}Sr_{0.175}NiO_2$ film deposited on $SrTiO_3$ single crystal and independent data fits to Eq. 1 (raw data reported by Fowlie *et al* [36] in their Figure S1,a [75]). For all fits $\rho_{sat} \to \infty$ (Eq. 1). Green balls indicate the bounds for which $\rho(T)$ data were fitted. Cyan balls indicate $T_{c,zero}$. The goodness of fit: (a) 0.99999; (b) 0.9998; (c) 0.99996; (d) 0.99998. The thickness of 95% confidence bands (pink shadow areas) is narrower than the width of the fitting lines.

**Figure 3.** $\rho(T)$ data for four pieces of the 7 nm thick $Nd_{0.825}Sr_{0.175}NiO_2$ film deposited on $SrTiO_3$ single crystal and global data fit to Eq. 2 (raw data reported by Fowlie *et al* [36] in their Figure S1,a [75]). For all fits $\rho_{sat} \to \infty$ (Eq. 2). Green balls indicate the bounds for which $\rho(T)$ data were fitted. Cyan balls indicate $T_{c,zero}$. Deduced parameters are $T_\omega = 442 \pm 5\ K$, $p = 2.07 \pm 0.3$. The goodness of fit: (a) 0.9998; (b) 0.9993; (c) 0.9987; (d) 0.99996. The thickness of 95% confidence bands (pink shadow areas) is narrower than the width of the fitting lines.

**Figure 4.** $\rho(T)$ data for four pieces of the 7 nm thick $Nd_{0.825}Sr_{0.175}NiO_2$ film deposited on $SrTiO_3$ single crystal and global data fit to Eq. 2 (raw data reported by Fowlie *et al* [36] in their Figure S1,a [75]) at $p = 5.0\ (fixed)$. Green balls indicate the bounds for which $\rho(T)$ data were fitted. Cyan balls indicate $T_{c,zero}$. Deduced Debye temperature is: $T_{\theta,global} = 313 \pm 1\ K$. For all fits $\rho_{sat} \to \infty$ (Eq. 2). The goodness of fit: (a) 0.9992; (b) 0.9995; (c) 0.9981; (d) 0.9997. The thickness of 95% confidence bands (pink shadow areas) is narrower than the width of the fitting lines.

**Figure 5.** $\rho(T)$ for six pieces of the 7.7 nm thick $Pr_{0.8}Sr_{0.2}NiO_2$ film deposited on $SrTiO_3$ single crystal and independent data fits to Eq. 1 (raw data reported by Fowlie *et al* [36] in their Figure S1,b [75]). For all fits $\rho_{sat} \to \infty$ (Eq. 1). Green balls indicate the bounds for which $\rho(T)$ data were fitted. Cyan balls indicate $T_{c,0.01}$. The goodness of fit: (a) 0.9998; (b) 0.99997; (c) 0.99992; (d) 0.99998; (e) 0.9998; (f) 0.99999. The thickness of 95% confidence bands (pink shadow areas) is narrower than the width of the fitting lines.

**Figure 6.** $\rho(T)$ for six pieces of the 7.7 nm thick $Pr_{0.8}Sr_{0.2}NiO_2$ film deposited on $SrTiO_3$ single crystal and global data fit to Eq. 2 (raw data reported by Fowlie *et al* [36] in their Figure S1,b [75]). For all fits $\rho_{sat,global} \to \infty$ (Eq. 2). Green balls indicate the bounds for which $\rho(T)$ data were fitted. Cyan balls indicate $T_{c,0.01}$. Deduced $T_{\omega,global} = 320 \pm 2\ K$, $p_{global} = 1.99 \pm 0.02$. The goodness of fit: (a) 0.9997; (b) 0.99990; (c) 0.9998; (d) 0.99998; (e) 0.9998; (f) 0.99996. The thickness of 95% confidence bands (pink shadow areas) is narrower than width of the fitting lines.

**Figure 7.** $\rho(T)$ data for six pieces of the 7.7 nm thick $Pr_{0.8}Sr_{0.2}NiO_2$ film deposited on $SrTiO_3$ single crystal and global data fit to Eq. 2 (raw data reported by Fowlie *et al* [36] in their Figure S1,a [75]) at $p = 5.0\ (fixed)$. For all fits $\rho_{sat} \rightarrow \infty$ (Eq. 2). Green balls indicate the bounds for which $\rho(T)$ data were fitted. Cyan balls indicate $T_{c,zero}$. Deduced Debye temperature is: $T_{\theta,global} = 241 \pm 1\ K$. The goodness of fit: (a) 0.9997; (b) 0.9996; (c) 0.9997; (d) 0.9998; (e) 0.9998; (f) 0.9997. The thickness of 95% confidence bands (pink shadow areas) is narrower than the width of the fitting lines.

**Figure 8.** $\rho(T)$ for eight pieces of the 7.7 nm thick $La_{0.8}Sr_{0.2}NiO_2$ film deposited on $SrTiO_3$ single crystal and independent data fits to Eq. 1 (raw data reported by Fowlie *et al* [36] in their Figure S1,c [75]). For all fits $\rho_{sat} \rightarrow \infty$ (Eq. 1). Green balls indicate the bounds for which $\rho(T)$ data were fitted. Cyan balls indicate $T_{c,0.01}$. The goodness of fit: (a) 0.99995; (b) 1.0; (c) 0.99998; (d) 0.99991; (e) 0.99995; (f) 0.99997; (g) 0.99994; (h) 1.0. The thickness of 95% confidence bands (pink shadow areas) is narrower than the fitting lines width.

**Figure 9.** $\rho(T)$ for eight pieces of the 7.7 nm thick $La_{0.8}Sr_{0.2}NiO_2$ film deposited on $SrTiO_3$ single crystal and global data fits to Eq. 2 (raw data reported by Fowlie *et al* [36] in their Figure S1,c [75]). For all fits $\rho_{sat} \rightarrow \infty$ (Eq. 1). Green balls indicate the bounds for which $\rho(T)$ data were fitted. Cyan balls indicate $T_{c,0.01}$. The goodness of fit: (a) 0.99994; (b) 0.99992; (c) 0.9989; (d) 0.9998; (e) 0.99991; (f) 0.99996; (g) 0.99998; (h) 0.99992. The thickness of 95% confidence bands (pink shadow areas) is narrower than the fitting lines width.

**Figure 10.** $\rho(T)$ for $Nd_{1-x}Sr_xNiO_2$ ($0.125 \leq x \leq 0.325$) films deposited on LSAT single crystal and data fits to Eq. 1 (raw data reported by Lee *et al* [73] in their Extended Data Figure 2 [73]). For all fits, if it is not specified, $\rho_{sat} \rightarrow \infty$ (Eq. 1). Green balls indicate the bounds for which $\rho(T)$ data were fitted. Cyan balls indicate $T_{c,0.1}$. The goodness of fit: (a) 0.99995; (b) 0.9998; (c) 0.9998; (d) 0.9998; (e) 0.99992; (f) 0.99992; (g) 0.99992; (h) 0.99992; (i) 0.99998; (j) 0.99994; (j) 0.99996. The thickness of 95% confidence bands (pink shadow areas) is narrower than the fitting lines width.

**Figure 11.** Deduced parameters for $Nd_{1-x}Sr_xNiO_2$ ($0.125 \leq x \leq 0.325$) films deposited on LSAT single crystal (raw data reported by Lee *et al* [73] in their Extended Data Figure 2 [73]).